\documentclass[aps,twocolumn,superscriptaddress,nofootinbib,groupedaddress,floatfix]{revtex4}

\usepackage{exscale}
\usepackage{graphicx}
\usepackage{amsmath}
\usepackage{latexsym}
\usepackage{amsfonts}
\usepackage{amssymb}
\usepackage{subfigure}

\def\be{\begin{equation}}
\def\ee{\end{equation}}
\def\bea{\begin{eqnarray}}
\def\eea{\end{eqnarray}}
\def\bma{\begin{mathletters}}
\def\ema{\end{mathletters}}

\def\0{\overline{0}}

\def\q0{\underline{0}}

\def\H{{\cal H}}

\def\L{{\cal L}}

\def\one{\leavevmode\hbox{\small1\normalsize\kern-.33em1}}

 \def\ket#1{|#1\rangle}

\newcommand{\Ket}[1]{\ensuremath{|#1\rangle}}

\newcommand\st[1]{{\mathrm{#1}}}

\newcommand{\entsp}{\stackrel{\scriptscriptstyle\wedge}{=}}

\newcommand{\eq}[1]{Eq.~(\ref{#1})}

\DeclareMathOperator{\Trace}{Tr}
\newcommand{\Tr}[1]{\ensuremath{\Trace\left(#1\right)}}

\newcounter{Anumctr}

\newcounter{Bnumctr}

\newcounter{Cnumctr}

\newcounter{alphnumctr}

\begin{document}

\title{Multipartite continuous-variable solution for the Byzantine agreement problem}

\author{Rodion Neigovzen}
\affiliation{Institut f\"ur Theoretische Physik, Universit\"at
Hannover, 30167 Hannover, Germany.}
\affiliation{Siemens AG, Corporate Technology, Otto-Hahn-Ring 6, D-80200 Munich, Germany.}

\author{Carles Rod\'o}
\affiliation{Grup de F\'isica Te\`orica, Universitat Aut\`onoma de
Barcelona, 08193 Bellaterra (Barcelona), Spain.}

\author{Gerardo Adesso}
\affiliation{Grup de F\'isica Te\`orica, Universitat Aut\`onoma de
Barcelona, 08193 Bellaterra (Barcelona), Spain.}
\affiliation{Dipartimento di Matematica e Informatica,
Universit\'a degli Studi di Salerno, Via Ponte Don Melillo, 84084
Fisciano (SA), Italy.}

\author{Anna Sanpera}
\affiliation{Grup de F\'isica Te\`orica, Universitat Aut\`onoma de
Barcelona, 08193 Bellaterra (Barcelona), Spain.}
\affiliation{Instituci\'o Catalana de Recerca i Estudis
Avan\c cats, Barcelona, 08021 Spain.}
\date{April 2, 2008}

\begin{abstract}
We demonstrate that the Byzantine Agreement (detectable broadcast) is
also solvable in the continuous-variable scenario with
multipartite entangled Gaussian states and Gaussian operations
(homodyne detection). Within this scheme we find that Byzantine
Agreement requires a minimum amount of entanglement in the
multipartite states used in order to achieve a solution. We
discuss realistic implementations of the protocol, which consider
the possibility of having inefficient homodyne detectors, not perfectly correlated outcomes, and
noise in the preparation of the resource states. The proposed protocol is proven to be robust and efficiently applicable under such non-ideal conditions.
\end{abstract}


\maketitle

\section{Introduction}
An algorithm is commonly defined as a set of rules for solving a
problem in a finite number of steps. One of the aims of quantum
information is to provide new protocols and algorithms which
exploit quantum resources to find a solution to problems which
either lack a solution using classical resources or the solution
is extremely hard to implement. The term ``Byzantine Agreement'' was
originally coined by Lamport \& Fischer \cite{lamport} in the
context of computer science to analyze the problem of fault
tolerance when a faulty processor is sending inconsistent
information to other processors. In a cryptographic context, it
refers to distributed protocols in which some of the participants
might have malicious intentions and could try to sabotage the
distributed protocol inducing the honest parties to take
contradictory actions between them. This problem is often
reformulated in terms of a Byzantine army where there is a
general commander who sends the order of attacking or retreating
to each one of his lieutenants. Those can also communicate
pairwise to reach a common decision concerning attacking or
retreating, knowing that there might be traitors among them
including the general commander. A traitor could create fake
messages to achieve that different parts of the army attack while
others retreat, which would put the army at a great disadvantage.
The question hence is whether there exists a protocol among all
the officials involved  that, after its termination, satisfies the
following conditions: The commanding general sends an order to his
$N-1$ lieutenants such that: (i) all loyal lieutenant obey the
same order (ii) if the commanding general is loyal, then every
loyal lieutenant obeys the order he sends. It is assumed that the
parties cannot share any previous setup. Lamport {\it et al.}
\cite{lamport2} proved that if the participants only shared
pairwise secure classical channels, then Byzantine Agreement or
broadcast is only possible if and only if $t<n/3$ where $n$ is the number of
players and $t$ is the number of traitors among them. In
~\cite{fitzi1} Fitzi and coworkers introduced a weaker
nevertheless important version of Byzantine Agreement known as
{\em detectable broadcast}. Detectable broadcast is said to be
achieved if the protocol satisfies the following conditions: (i)
if no player is corrupted, then the protocol achieves broadcast
and (ii) if one or more players are corrupted, then either the
protocol achieves Byzantine Agreement or all honest players abort
the protocol. Thus, in a detectable broadcast protocol, cheaters
can force the protocol to abort, i.e. no action is taken. In such
cases all honest players agree on aborting the protocol so that
contradictory actions between the honest players are avoided. In
the same paper \cite{fitzi1}, Fitzi, Gisin and Maurer
\cite{fitzi1} devised a solution to the detectable broadcast
problem using multipartite entanglement as a quantum resource.
Later on Fitzi {\it et al.} \cite{fitzi2} and Iblisdir and Gisin
\cite{iblisdir2004} showed that a Quantum Key Distribution (QKD)
protocol, which guarantees a private sequences of classical data
shared between pairs of parties, suffices to solve detectable
broadcast. This situation is reminiscent of quantum cryptography,
in which two seemingly independent main schemes exist for QKD, the
prepare-and-measure Bennett and Brassart 1984 (BB84) scheme \cite{BB84} which does not use
entangled states shared between Alice and Bob, and the Ekert 1991 (Ekert91)
scheme \cite{Ekert91} where indeed entanglement is explicitly
distributed and the security is guaranteed by Bell's theorem.
However, the two schemes have been shown to be completely
equivalent \cite{Bennett92}, and specifically entanglement stands
as a precondition for any secure key distribution \cite{curty}.

Detectable broadcast might be regarded in a similar view as
quantum key distribution and, arguably, the same reasoning applies
to the different protocols advanced for its solution, making
explicit or implicit use of multipartite entanglement. In this
paper, we adopt an approach of the Ekert91, guided by the physical
motivation of studying the performance and the usefulness of
multipartite entangled states as operational resources to achieve
detectable broadcast. While protocols for this task exist for
qutrits \cite{fitzi1} and qubits \cite{cabello03,cabello07}, in
this paper we investigate the possibility of solving detectable
broadcast with continuous-variable (CV) systems
\cite{brareview,covaqial}, namely with Gaussian states and
performing Gaussian operations only. The motivations for this
approach are manifold. On the practical side, the recent
progress in CV QKD \cite{grosschapter} has shown that the use of
efficient homodyne detectors, compared to photon counters employed
in BB84 schemes, enables the distribution of secret keys at faster
rates over increasingly long distances \cite{cvqkdexp}. On the
theoretical side, multipartite entanglement is a central concept
whose understanding and characterization  (especially in
high-dimensional and CV systems), despite recent efforts, is far
from being complete. It is important, therefore, to approach this
task {\em operationally}, i.e. by connecting entanglement to the
success or to the performance of diverse quantum information and
communication tasks \cite{PlenioVirmani}, while exploiting the
differences between discrete-variable and CV scenarios. For
Gaussian states of light fields, which are presently the
theoretical and experimental pillars of quantum information with
CVs \cite{brareview,covaqial,myreview}, an important result in
this respect is due to van Loock and Braunstein \cite{network}.
They introduced a scheme to produce fully symmetric
(permutation-invariant) $n$-mode Gaussian states exhibiting
genuine multipartite entanglement. Furthermore, they devised a
communication protocol, the ``quantum teleportation network,''
for the distribution of quantum states exploiting such resources, which has been
experimentally demonstrated for $n=3$ \cite{naturusawa}. The
optimal fidelity characterizing the performance of such a protocol
yields an operational quantification of genuine multipartite
entanglement in symmetric Gaussian states. This quantification is
equivalent to the information-theoretic ``residual contangle''
measure, a Gaussian entanglement monotone, emerging from the
monogamy of quantum correlations \cite{telestrong,contangle}.
Other applications of multipartite Gaussian entanglement have been
advanced and demonstrated, and the reader may find more details in
\cite{brareview,covaqial}, as well as in the two more recent
complementary reviews \cite{myreview} (theoretical) and
\cite{njprev} (experimental). In this respect, the usage of
multipartite entangled Gaussian resources for the Byzantine
Agreement problem was, to the best of our knowledge, not
previously investigated.

Here we propose a protocol to solve detectable broadcast
with symmetric multimode entangled Gaussian states and homodyne detection.
This provides an alternative interpretation of multipartite
Gaussian entanglement as a resource enabling this kind of secure
communication. We concentrate on the case of three parties, and
remarkably find that not all three-mode symmetric entangled Gaussian states
are useful to achieve a solution: To solve detectable broadcast
there is a minimum threshold in the multipartite entanglement.
This is at variance with the two-party QKD counterpart: In there,
all two-mode entangled Gaussian states are useful to obtain a
secure key using Gaussian operations \cite{navascues}. We
eventually discuss how our protocol can be implemented in
realistic conditions, namely considering detectors with finite efficiency and  yielding not
perfectly matched measurement outcomes, and Gaussian resources
which are not ideally pure, but possibly (as it is in reality)
affected by a certain amount of thermal noise. We show that under
these premises the protocol is still efficiently applicable to
provide a robust solution to detectable broadcast over a broad
range of the involved parameters (noise, entanglement, measurement
outcomes, and uncertainties), paving the way towards a possible experimental
demonstration in a quantum optical setting.

The paper is organized as follows. In Sec. II we outline the
steps needed to solve detectable broadcast in the discrete
scenario following \cite{fitzi1,cabello03}. In Sec. III, we
first review briefly the basic tools needed to describe
multipartite Gaussian states and measurements. Building on the
ideas presented in Section II we construct a protocol adapted to
the continuous-variable case. In Sec. IV we discuss security on
the distribution of quantum states for the protocol. Section V is
devoted to analyze the effects of malicious manipulations of the
data by dishonest parties and its detection by the honest ones. In
Sec. VI we focus on the efficiency of the proposed protocol and
we extend our results to a realistic practical scenario, relaxing
the conditions for obtaining of correlated outputs between the
players and assuming noise in the preparation of the input states.
Finally, in Sec. VII we present our concluding remarks.

\section{Detectable broadcast protocol}

The protocols to solve detectable broadcast using entanglement as a
resource are based on three differentiated steps:
\begin{enumerate}
  \item[(i)] Distribution of the quantum states,
  \item[(ii)] Test of the distributed states,
  \item[(iii)] Protocol by itself.
\end{enumerate}
Step (iii) is fundamentally classical, since it uses the
outputs of the different measurements of the quantum states to
simulate a particular random generator (``primitive''). Let us
consider the simplest case, in which only three parties are involved.
They are traditionally denoted by $S$ (the sender, i.e. commander
general) and the receivers $R_0$ and $R_1$, and at most, only one
is a traitor. In this case, the primitive generates for every
invocation a random permutation of the elements $\{0,1,2\}$ with
{\em uniform distribution}, i.e. $(t_S, t_{R_0}, t_{R_1}) \in
\{(0,1,2),(0,2,1),(1,0,2),(1,2,0),(2,0,1), (2, 1, 0)\}$. In this
primitive, no single player $n$ can learn more about the
permutation than the value $t_n$ ($n=\{S, R_0, R_1\}$) which
she or he, receives i.e., each player ignores how the other two values
are assigned to the other players. Furthermore, nobody else
(besides the parties) have access to the sequences.

Entanglement is used in the protocol to distribute classical private
random variables with a specific correlation between the players, in
such a way that any malicious manipulation of the data can be
detected by all honest parties allowing them to abort the protocol.
In the discrete variable case such a primitive can be implemented
with qutrits using e.g. Aharonov states $\Ket{\mathcal{A}} =
\frac{1}{\sqrt{6}}( \Ket{0,1,2} + \Ket{1,2,0} +
\Ket{2,0,1}-\Ket{0,2,1} - \Ket{1,0,2} - \Ket{2,1,0})$. This choice
allows the distribution and test part to be secure (for details
see ~\cite{fitzi1}). Whenever the three qutrits are all measured
in the same basis,  all three results are different. Hence --
after discarding all the states used for the testing of the
distributed states [step (ii)] -- the players are left with a
sequence of $L$ outputs that reproduces the desired primitive. We
schematically represent this primitive by the table below.
\begin{center}
 \begin{tabular}{|c|c|c|c|c|c|c|c|c|c|c|}
 \hline $j$   & 1 & 2 & 3 & 4 & 5 & 6 & 7 & 8 & 9 & $\ldots$\\
 \hline
 \hline $S$   & 2 & 0 & 0 & 1 & 2 & 1 & 0 & 2 & 1 & $\ldots$\\
 \hline $R_0$ & 1 & 1 & 2 & 0 & 1 & 0 & 1 & 0 & 2 & $\ldots$\\
 \hline $R_1$ & 0 & 2 & 1 & 2 & 0 & 2 & 2 & 1 & 0 & $\ldots$\\
 \hline
 \end{tabular}
\end{center}
After accomplishing the distribution and test part of the protocol
[steps (i) and (ii)], the sender $S$ will broadcast a bit $b \in
\{0,1\}$ ($0\entsp$``attack'', $1\entsp$``retreat'') to the two
receivers using classical secure channels.

Following \cite{fitzi1} the broadcast [step (iii)] proceeds as
follows:

\noindent (iii-1): We denote by $b_i$ the bits received by $R_i$,
$i=0,1$ (notice that if the sender is malicious, the broadcasts
bits $b_i$ could be different). Each receiver $R_i$ demands to $S$
to send him the indices $j$ for which $S$ obtained the result $b_i$.
Each player $R_i$ receives a set of indices $J_i$.\\
\noindent (iii-2): Each $R_i$ test consistency of his own data,
i.e. checks weather his output on the set of indices he receives
($J_i$) are  all of them different from $b_i$. If so the data is
consistent and he settles his flag to $c_i=b_i$. otherwise his
flag is settled to $c_i=\perp$.\\
\noindent (iii-3): $R_0$ and $R_1$ send their flags to each other.
If both flags agree, the protocol terminates with all honest
participants agreeing on $b$.\\
\noindent (iii-4): If $c_i=\perp$, then player $R_i$ knows that
$S$ is dishonest, the other player is honest and accepts his flag.\\
\noindent (iii-5): If both $R_0$ and $R_1$ claim to have
consistent data but $c_0\neq c_1$, player $R_1$ demands from $R_0$
to send him all the indices $k\in J_0$ for which $R_0$ has the
results $1-c_0$. $R_1$ checks now that (i) all indices $k$ from
$R_0$ are not in $J_1$ and (ii) the output $R_1$ obtains from
indices $k$ correspond to the value 2. If this is the case, $R_1$
concludes that $R_0$ is honest and changes his flag to $c_0$. If
not, $R_1$ knows that $R_0$ is dishonest and  he keeps his flag to
$c_1$. Detectable broadcast is in this way achieved.\\

\section{Continuous-variable primitive}\label{secProt}

Before describing in detail how to achieve the primitive presented
in the above section in the CV set up, let us briefly review the
most basic ingredients needed to describe Gaussian states
\cite{brareview,myreview}. We consider here quantum systems of $n$
canonical degrees of freedom, often called modes, associated to a
Hilbert space $\H=\L^2(\Re^{2n})$. The commutation relations
fulfilled by the canonical coordinates $\hat{R}^T =
(\hat{X}_1,\hat{P}_1,\ldots,\hat{X}_n,\hat{P}_n)=
(\hat{R}_1\ldots,\hat{R}_{2n})$ are represented in matrix form
by the symplectic matrix $J$: $[\hat{R}_a,\hat{R}_b] =
i(J_n)_{ab}$, with $a,b=1,\ldots,2n$, and
\begin{equation}
 J_n=\oplus_{i=1}^n J \quad\quad J\equiv
 \begin{pmatrix}
 0 & 1 \cr
 -1 & 0
 \end{pmatrix} .
\end{equation}
Gaussian states are, by definition, completely described by their
first and second moments. Hence, rather than describing them by
their (infinite-dimensional) density matrix $\hat\rho$, one can
use the Wigner function representation
\begin{equation}\label{wignerfunc}
 W_{\rho}\left(\zeta\right)=\frac{1}{\pi^{n}\sqrt{\det{\gamma}}}
 \exp{\left[-(\zeta-d)^\st{T}\gamma^{-1}(\zeta-d)\right]}
\end{equation}
which is a function of the first moments through the displacement
vector $d$ (a $2n$ real vector), and of the second moments through
the covariance matrix $\gamma$ (a $2n\times 2n$ symmetric real
matrix). Here $\zeta \in \mathbb{R}^{2n}$ is a phase-space vector.
The positivity condition of $\hat\rho$ implies that
$\gamma+iJ_n\geq 0$.

Following the discussion of the protocol [step (iii)] for the
discrete case, we choose to use as a resource a pure, fully
inseparable tripartite Gaussian state $\hat\rho_a$, completely
symmetric under the interchange of the modes \cite{network}, whose
covariance matrix $\gamma(a)$ accepts the following
parametrization \cite{giedke}:
\begin{equation}\label{e:gamma}
 \gamma(a) =\left(
 \begin{array}{cccccc}
 a&0&c&0&c&0 \\ 0&b&0&-c&0&-c \\ c&0&a&0&c&0 \\ 0&-c&0&b&0&-c \\
 c&0&c&0&a&0 \\ 0&-c&0&-c&0&b
 \end{array}\right),
\end{equation}
with $a \ge 1$ and
\begin{eqnarray}
 b=\frac{1}{4}(5a-\sqrt{9a^2-8}),\\
 c=\frac{1}{4}(a-\sqrt{9a^2-8}).
\end{eqnarray}
The condition for full inseparability (truly multipartite entanglement) across the
$1\times1\times1$ mode partition comes from the nonpositivity of the
partial transpose (NPT-criteria) and reads \cite{giedke}:
\begin{eqnarray*}
 \gamma+iJ_A&\not\geq&0\\
 \gamma+iJ_B&\not\geq&0\\
 \gamma+iJ_C&\not\geq&0
\end{eqnarray*}
with $J_A=J^\st{T}\oplus J\oplus J$, $J_B=J\oplus J^\st{T}\oplus
J$, $J_C=J\oplus J\oplus J^\st{T}$. It follows that $\gamma(a)$ is
fully inseparable as soon as $a>1$. Quantitatively, the genuine
tripartite entanglement of the states of \eq{e:gamma}, as measured
by the residual contangle \cite{contangle}, is a monotonically
increasing function of $a$ and diverges for $a \rightarrow
\infty$.

So far $\gamma(a)$ appears as the ``equivalent'' CV version of the
discrete Aharonov state  $\ket {\cal A} $, i.e. fully inseparable
and completely symmetric under exchange of the three players. One is
tempted to infer, therefore, that the discussed primitive of the
discrete case can be straightforwardly generalized to the
continuous one. A standard way of transforming the correlations of
the shared entangled quantum states $\hat{\rho}_a$ into a sequence
of classically correlated data between the 3 players is to perform
a homodyne measurement of the quadratures of each mode. Denoting
by $\hat{X}_{S}$, $\hat{X}_{R_0}$, $\hat{X}_{R_1}$ the  position
(or momentum) operator of each mode, and by ${x}_{S}$,
${x}_{R_0}$, ${x}_{R_1}$ the output of the respective measurement,
the players can communicate classically with each other and agree,
for instance, only on those values for which either
$\left|{x}_{S}\right|=\left|{x}_{R_0}\right|=\left|{x}_{R_1}
\right|=\{0,x_0\}$ with $x_0>0$. In this way each player can
associate the logical trit ($t=0,1,2$) to a null, positive, or
negative result, respectively. The probability distribution that
measuring the quadratures $\hat{X}_{S}$, $\hat{X}_{R_0}$,
$\hat{X}_{R_1}$ produces an outcome $t_{S}=j$, $t_{R_0}=k$,
$t_{R_1}=l$, with $(j,k,l)\in \{0,1,2\}$ and uncertainty $\sigma$
is given by:
\begin{eqnarray}\label{overlap}
 {\cal P}(j,k,l)&=&\Tr{\hat{\rho}_M^{\{j,k,l\}}\hat{\rho}_a}\\
 &=&(2\pi)^n\int W_M^{\{j,k,l\}}(\xi)W_a(\xi)d^{2n}\xi\nonumber.
\end{eqnarray}
Here $\hat{\rho}_{M}^{j,k,l} = \hat{\rho}_S^{t_S=j} \otimes
\hat{\rho}_{R_0}^{t_{R_0}=k} \otimes \hat{\rho}_{R_1}^{t_{R_1}=l}$
describes the separable state of the 3 modes obtained after each
party has measured its corresponding quadrature and obtained an
output $j,k,l$. Thus, the state of the system after the measure
can be described by a fully separable covariance matrix
$\gamma_M$,
\begin{equation}
 \gamma_M =\left(
 \begin{array}{ccc}
 \gamma_m&0&0\\
 0&\gamma_m&0\\
 0&0&\gamma_m
 \end{array}\right).
\end{equation}
where each party has a pure one-mode Gaussian state with covariance
matrix
\begin{equation}
 \gamma_m =\left(
 \begin{array}{cc}
 \sigma^2&0\\
 0&1/\sigma^2
 \end{array}\right)
\end{equation}
and displacement vector $d_m^T=(-|x_0|,0)$. On the other hand  $W_a$ is the
Wigner function of the initial tripartite entangled state
$\hat{\rho}_a$ described by $\gamma(a)$ and $d$.

Since Gaussian states are symmetric with respect to their
displacement vector, it is easy to see that a mapping into
classical trits such that all possible outputs $\{0,1,2\}$ occur
with the same probability is not possible. Therefore, the
primitive discussed in the discrete case must be modified
when adapted to the CV scenario. The way we find to
overcome this asymmetry of the output's probabilities relies on
joining the correlations from pairs of quantum states. First we
map the quantum correlations involved in each single quantum state
just to classical bits  by a ``sign binning'' (as in CV QKD \cite{navascues}).
That is, first we keep only the results for which the three
players obtain a coincident output $|x_S|=|x_{R_0}|=|x_{R_1}|=
x_0>0$, and associate the logical bit $b_n=0 (1)$, where $n=\{S,R_0,R_1\}$,
to positive (negative) value of the coincident quadrature $x_0 (-x_0)$. In every measurement the sender makes public his outcome result $|x_0|$, whatever the output is, but not its sign in such a way that in principle all states are going to be used in the protocol.
Second, we construct a primitive consisting
of a random permutation of the elements $(i,j,k) \in
\{(0,1,1),(1,0,1),(1,1,0)\}$. Compulsory for the implementation of
the primitive is that every element of the primitive appears with
equal probability and any other combination of the outputs not
regarded by the primitive is exceedingly small compared to the
allowed permutations. In other words, we demand that $p \equiv {\cal
P}(0,1,1)={\cal P}(1,0,1)={\cal P}(1,1,0)$, and, denoting by
$\delta \equiv {\cal P}(else)$, we require that the corresponding
conditional probabilities,
$$\tilde{{\cal P}}(b_S,b_{R_0},b_{R_1})=\frac{{\cal
P}(b_S,b_{R_0},b_{R_1})}{\sum_{i,j,k=\{0,1\}} {\cal P}(j, k,
l)}\,,$$ fulfill $\tilde p \rightarrow\frac{1}{3}$ and
$\tilde{\delta}\rightarrow0$.

\begin{figure}[t]
 \centering
 \includegraphics[width=7cm]{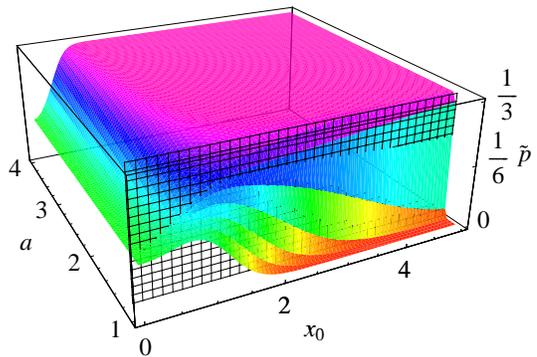}
 \caption{(Color online) Plot of the conditional probability
 $\tilde{p}$ as a function of the measurement outcome $x_0$ and
 the shared entanglement $a$, for pure symmetric tripartite
 Gaussian resource states. Detectable broadcast is ideally solvable in
 the huge, unbounded region of $x_0 \gg 0$, $a \gg 1$, where
 $\tilde{p} \rightarrow 1/3$. The entanglement threshold
 $a=a_{thresh}\equiv 5\sqrt2/6$ is depicted as well (wireframe
 surface). All of the quantities plotted are dimensionless.}\label{paramideals}
\end{figure}

If the above conditions are met, it is possible then by invoking
two consecutive times the previous generator [i.e. using a pair of
quantum states of the class \eq{e:gamma}], to map the bit values
to trit-values (0,1,2) plus an additional undesired element
``$u$''. For instance, the players, after keeping only those bits
obtained by coincident quadrature outputs  $\pm x_0$, use two
consecutive bits $m$ and $m+1$ for the following association:
\begin{eqnarray}
 \nonumber (1,0)\rightarrow {\bf 0}\\
 \nonumber (0,1)\rightarrow {\bf 1}\\
 \nonumber (1,1)\rightarrow {\bf 2}\\
 \nonumber (0,0)\rightarrow {\boldsymbol{u}}\\\nonumber
\end{eqnarray}
Thus, by concatenating two invocations, one generates the permutations
corresponding to the primitive
$\{(0,1,2),(0,2,1),(1,0,2),(1,2,0),(2,0,1),(2,1,0)\}$ plus a
permutation of undesired elements $(u,2,2)$, $(2,u,2)$, $(2,2,u)$
which will be discarded during the protocol. With all these tools
at hand, optimal results for the desired probabilities are
achieved for a displacement vector of the form
$d^T=-\frac{x_0}{3}(1,0,1,0,1,0)$, and yield:
\begin{eqnarray}\label{eq:pAbs}
 \delta_{1}&=&{\cal P}(1,1,1)=C(a,\sigma)\exp\left(-\frac{4}{3}
 \frac{x_0^2}{K_1}\right),\nonumber\\
 \delta_{2}&=&{\cal P}(0,0,0)=C(a,\sigma)\exp\left(-\frac{16}{3}
 \frac{x_0^2}{K_1}\right),\\
 \delta_{3}&=&{\cal P}(0,0,1)={\cal P}(0,1,0)={\cal P}(1,0,0)\nonumber\\
 &=&C(a,\sigma)\times\exp\left[ \frac{-4 x_0^2\left(\sigma^2+
 \frac{1}{4}\left[5a-\sqrt{9a^2-8}\right]\right)}{K_1 K_2}\right],\nonumber
\end{eqnarray}
and
\begin{equation}
 \begin{split}
 p\ &=\ {\cal P}(0,1,1)={\cal P}(1,0,1)={\cal P}(1,1,0)\\
 &=\ C(a,\sigma)\exp\left(-\frac{8}{3}\frac{x_0^2}{K_2}\right),
\end{split}
\end{equation}
with coefficients $K_1=\sigma^2+\frac{1}{2}\left[3a-\sqrt{9a^2-8}
\right]$, $K_2=\sigma^2+\frac{1}{4}\left[3a+\sqrt{9a^2-8}
\right]$, and prefactor \[\begin{split}C&(a,\sigma)=\left[
\det{\left( \frac{\gamma_{M}+\gamma(a)}{2}\right) }\right]
^{-\frac{1}{2}}\\=&\frac{8}{\left(a-c+\sigma^2 \right) \left(
b+c+\frac{1}{\sigma^2}\right) \sqrt{\left(a+2c+\sigma^2 \right)
\left( b-2c+\frac{1}{\sigma^2}\right)}}.\end{split}\] The
probability distribution depends on the parameters $a$, $x_0$ and
$\sigma$. In the case $a\gg1$, it can be seen that the conditional
probabilities satisfy the following:
\begin{equation}\label{eq:approxG_a}
 \tilde{p}=\frac{1}{3}-\frac{4}{9}k+O(k^2);\;\;\;
 \tilde{\delta}_i\rightarrow 0
\end{equation}
with $k=\exp{\left[-\frac{4}{3}(\frac{x_0}{\sigma})^2\right]}$.
In the limit of a perfect, zero-uncertainty homodyne detection
($\sigma \rightarrow 0$), $\tilde{p}$ converges exactly to $1/3$.
In general, there exists a large region in the parameters space
for which $\tilde{p}\rightarrow \frac{1}{3}$ and $\tilde
\delta_i\rightarrow 0$, as depicted in Fig.~\ref{paramideals}.

However, there is a {\em lower} bound on the entanglement content
of the symmetric Gaussian states of \eq{e:gamma} in order to
fulfill the above conditions. Only for $a\geq a_{thresh} \equiv
\frac{5\sqrt{2}}{6} \approx 1.18$, one has that $\tilde {p} >
\tilde{\delta_{i}}$, which is a necessary condition to implement
the primitive. This indicates that not all pure three-mode symmetric
full entangled Gaussian states can be successfully employed to
solve detectable broadcast via our protocol. This
entanglement threshold is an {\em a priori} bound which does not
depend on the specific form of the employed resource states. For
any parametrization of the covariance matrix of $\hat{\rho}_a$,
which is obtainable from $\gamma(a)$ by local unitary squeezing
transformations (hence at fixed amount of tripartite entanglement),
the same condition discriminating useful resource states is
analytically recovered. We remark that this bound becomes tight in
the limit $x_0 \rightarrow \infty$, meaning that $a \geq
a_{thresh}$ becomes necessary {\em and} sufficient for the
successful implementation of the primitive.

\section{Distribution and test}

We move now to the distribution \& test part of the protocol which
represents the first step in the execution of the appropriate
primitive and has only two possible outputs: Global success or
global failure. In the case of failure, a player assumes that
something went wrong during the execution of the protocol and aborts
any further action. In the case of global success, each of the
parties ends up with a set  $\{K\}$  of classical data, and the
protocol proceeds classically, according to the steps explained in
Section II.

From now on we assume that the players share pairwise secure
classical channels and secure (noiseless) quantum channels. The
secure distribution \& test part uses correlations to validate the
fairness of the other parties. Therefore, quantum states are sent
through noiseless quantum channels and measures are performed
massively. In doing so, it is possible to detect manipulation of the
data on a statistical basis and abort the protocol if necessary. In
this Section we mostly focus on the issue of security in the
distribution and security test of the data. This should permit the
detection of any malicious manipulation of the data. The explicit
effects of sabotage actions will be reported in Section V. We
postpone the important issue of how well the protocol will succeed
in the case that the outputs of the measurements are not perfectly
correlated to Section VI. The reader not interested in the detailed
description of the distribution \& test part can skip the rest of
this section and jump to Section V or VI instead.

The distribution \& test part proceeds as follows:\\
\noindent (i-1) Without losing generality, let us assume that
$R_1$ prepares a large number, $M$, of tripartite systems in state
$\hat{\rho}_a$ (i.e. with covariance matrix $\gamma(a)$ and displacement
$d$) and sends one subsystem to $S$ and another to $R_0$.\\
\noindent (i-2) $R_1$ wants to check if the distribution of states
is faithfully achieved. To this aim she or he chooses randomly a set of
indices for player $S$, $\{K_S\}$, and a disjoint set of indices for
player $R_0$, $\{K_{R_0}\}$, and sends these two sets over secure
classical channels to the corresponding players. Player $S$ sends
his or her $\{K_S\}$ subsystems to player $R_0$. For each $m\in
\{K_S\}$, $R_0$ measures the two subsystems in his or her possession
and $R_1$ measures his subsystem. After communication of their
results over secure classical channels, they agree on those indices
$\{\tilde K_S\}\subseteq K_S$ for which $|x_{R_0}|=|x_{R_1}|=x_0$.
$R_1$ and $R_0$ check now whether the correlations predicted by the
primitive occur: $\tilde p=\tilde{\cal P}(011)=\tilde{\cal P}(101)=\tilde{\cal
P}(110)=\frac{1}{3}$, $\tilde \delta=0$. If the test
was successful, i.e. if the measurement results were consistent with
the assumption that the states have been distributed correctly, the
players $i\in \{R_0, R_1\}$ set the flag $f_{i}=1$, otherwise
$f_{i}=0$.
In an analogous way, the test is performed for $S$.\\
\noindent (i-3) Players $S$, $R_0$ and $R_1$ send their flags to
each other. Every player who receives a flag ``$0$'', sets his
flag also to ``$0$''. Every player with flag ``$0$'' aborts the
protocol. Otherwise, the execution of the protocol proceeds. This
step terminates the distribution \& test of the quantum systems.

In the second phase of the protocol a selection of the distributed
systems is chosen to establish the bit sequences which will be used
to implement the quantum primitive. In this phase, again honest
parties may abort
the protocol if malicious manipulations occur.\\
\noindent (ii-1) The players $S$, $R_0$ and $R_1$ agree upon a set
of systems
which have not been discarded during the distribution \& test part.\\
\noindent (ii-2) Player $S$ chooses (randomly) two disjoint sets
of subsystems labeled by indices $L_S^i\subset\tilde{M}$. He or she sends
the set $L_S^i$ to player $i$ and demands player $i$ to send via a
noiseless quantum channel his or her subsystems $m\in L_S^i$ to
him or her. In each case, the (random) choice $L_S^i$ is secret to
party $j$, i.e. player $R_1$  has no information whatsoever about
the set $L_S^{R_0}$. An analogous procedure is adopted by $R_0$ and
$R_1$.\\ \noindent (ii-3) After measuring their whole sequence of
subsystems, player $i\in\{S,R_0,R_1\}$ announces publicly, the set
of indices $\{\hat{M}^m_i\}\in\hat{M}_{i}$ for which the output of
the quadrature measurement was $|x_0|$ The order in which the
players announce their measurement results can be specified
initially and based, e.g. on a rotation principle. (Notice that the
announcement of $\left|x_i\right|=x_0$ during the actual
measurement phase would make possible an effective traitor
strategy, since he could manipulate combinations on a systematic
basis).\\ \noindent (ii-4) Without loss of generality, let us
explicitly describe this step of protocol for player $S$. From the
following sets $L_{S}^{R_0}\cap\hat{M}^{m}_{R_1}=:U^{R_1}_S$ and
$L_S^{R_1}\cap\hat{M}^{m}_{R_0}=:U^{R_0}_S $ let
$\tilde{U}^{R_i}_{S}\subseteq U^{R_i}_{S}$ for $i\in\{0,1\}$ be
the index set for which the player $S$ measured $\pm x_0$ twice.
Analogously to the first phase of the protocol, player $S$ can
test if the outputs of his measures agree with the correlations of
a proper  primitive ($p(00)=p(10)=p(01)=1/3$ and $p(11)=0$) If
the test is successful, the player sets his flag $f_{S}=1$,
otherwise $f_{S}=0$. The same procedure is analogously performed by
$R_0$ and $R_1$.
From this step on, the players deal exclusively with the outputs of
their measures,
i.e. classical data and secure classical channels.\\
\noindent (ii-5) $S$ checks correlation on his outputs in the
following set
$\hat{M}:=\hat{M}^{m}_{S}\cap\hat{M}^{m}_{R_0}\cap\hat{M}^{m}_{R_1}
\subseteq\hat{M}^{m}_{S}$.
If the test is successful, $S$ sets his flag $f_{S}=1$, otherwise
$f_{S}=0$.
$R_0$ and $R_1$ do the equivalent step.\\
\noindent (ii-6): From a randomly chosen set $V^{S}\subset\hat{M}$
player $S$ demands from $R_0$ and $R_1$ their measurement results.
$S$ tests this control sample for the assumed primitive. If
the test is successful, $S$ sets his flag $f_{S}=1$, otherwise
$f_{S}=0$. $R_0$ ($R_1$) perform this step with a set
$V^{R_0}\subset\hat{M}\backslash V^{S}$
($V^{R_1}\subset\hat{M}\backslash (V^{S}\cup V^{R_0})$) respectively.\\
\noindent (ii-7) Every player with a flag $0$ aborts the execution
of the protocol. Otherwise the players agree upon a set
$W:=\hat{M}\backslash(V^{S}\cup V^{R_0}\cup V^{R_1})$ as the result
of the invocation of the primitive, which consists in an even number
of elements. This step concludes the distribution part of the
protocol.

\section{Error Estimation}

We examine here two possible sources of error, which can
occur during the implementation of the primitive with Gaussian
states. While the first source is inherent to the system, the
second is caused by an active adversary intervention of a
participating party. The inherent error is caused by the
occurrence of non consistent combinations on the invocation of
the primitive, that is, the occurrence of outputs $(1,1,1)$,
$(0,0,0)$, $(0,0,1)$, $(0,1,0)$ and $(1,0,0)$ has to be
considered. This error will propagate along the protocol, so that
the probability of finding a combination which is not appropriate
is bounded from above from $\eta=1-(3\tilde p)^2$.

The second source of errors we want to discuss here corresponds to
the local actions that one of the players could do in order to
manipulate the measurement results of other players. For instance,
let us assume that the player $R_0$ has malicious intentions and
wants to shift the local component of the displacement vector of the
distributed state using local transformations. Note that a
shift of the quadrature output $x_0$, provides the same error in the
probability distribution of the outputs as a shift in the
corresponding displacement vector. In other words, both types of
manipulations produce the same change in the probabilities as
calculated in Eq.(\ref{overlap}). Parameterizing the shift in the
displacement vector by the parameter $K$, $d^T \mapsto {(d')^T}
= -\frac{x_0}{3}(1,0,1,0,K,0)$, it is interesting to see how that
affects the conditional probabilities $\tilde p$ and $\tilde
\delta$. If probabilities were changed, player $R_0$ could try to
determine [via subsequent communication with the other players (step
ii-3)], with certain probability, the occurrence of the outputs of the
other players, thereby gaining additional information. Notice also
that $S$ and $R_1$ cannot realize the local manipulation of $R_0$
without classical communication between them. This can be trivially
seen by realizing that the partial trace ${\rm Tr}_{R_0}
(\hat \rho_a(\gamma, d'))=\int W'_{\xi}dx_{R_0}dp_{R_0}=
W'_{\xi}(\gamma_{S,R_1},d_{S,R_1})={\rm Tr}_{R_0}(\hat \rho_a(\gamma,d))$
with
\begin{equation}
 \gamma_{S,R_1}=\left( \begin{array}{cccc}
 a & 0 & c & 0 \\
 0 & b & 0 & -c \\
 c & 0 & a & 0 \\
 0 & -c & 0 & b
 \end{array} \right)\quad\textnormal{and}\quad
 d_{S,R_1}=\left( \begin{array}{c}
 x_1 \\
 0 \\
 x_1 \\
 0
 \end{array} \right).
\end{equation}
Thus the most plausible strategy for a traitor could consist of the
following: (i) Discrediting honest players by manipulating the
displacement vector in such a way that non consistent combinations
appear, (ii) hide successful measurements to the honest players
which result in combinations that might be disadvantageous. It is
tedious but straightforward to show that by making use of the test
steps (ii-4)--(ii-6) honest parties can detect the effects of such
manipulations.

\section{Efficient realistic implementation}\label{secEff}

The protocol, as discussed in the previous Sections, constitutes a
nice proof-of-principle of the fact that detectable broadcast is
solvable in the CV scenario using multipartite Gaussian
entanglement. However, it suffers from its reliance on two main
idealizations which render the practical implementation of the
primitive unrealistic, or, better said, endowed with zero
efficiency. Specifically, we have requested that (i) a {\em pure}
tripartite symmetric Gaussian state is distributed as the entangled
resource; and (ii) when the three parties measure (via homodyne
detection) the position of their respective modes, their measurement is taken to be ideal, that is not affected by any uncertainty, and moreover all parties must
obtain, up to a sign, {\em the same} outcome $x_0$. In reality,
assumption (i) is unjustified as inevitable imperfections and losses
result instead in the production of mixed, thermalized states; on
the other hand, the probability associated to (ii), and hence the
probability of achieving broadcast, is vanishingly small \cite{carles}. It is
interesting, in view of potential practical implementations of our
scheme, to study here how its success is affected, and possibly
guaranteed, by relaxing the above two assumptions.

To deal with (i), let us recall that the tripartite entangled
Gaussian states of \eq{e:gamma} can be produced in principle by
letting three independently squeezed beams (one in momentum, and two
in position) interfere at a double beam splitter, or ``tritter''
\cite{branature}, as proposed by van Loock and Braunstein
\cite{network}. In practice, the parametric nonlinear process
employed to squeeze the vacuum is affected by losses which result in
the actual generation of squeezed {\em thermal} states in each
single mode. Before the tritter, one then has three independent
Gaussian modes with covariance matrices $\gamma_{1}^{{\rm in}}(s,n)={\rm
diag}\{n s,\,n/s\}$, $\gamma_{2}^{{\rm in}}(s,n)=\gamma_{3}^{{\rm in}}(s,n)={\rm
diag}\{n/s,\,n s\}$, respectively, where $s=\exp(2r)$ (with $|r|$ the squeezing degree in each single mode) and $n\ge 1$ is the noise parameter
affecting each mode. The noise $n$ is related to the initial
marginal purity $\mu_k^{{\rm in}}$ of each single mode by
$n=1/\mu_k^{{\rm in}}$ \cite{notepurity}, and corresponds to a mean
number of thermal photons given by $\bar n^{{\rm th}}=(n-1)/2$. For $n=1$,
each mode is in the ideally pure squeezed vacuum state. After the
tritter operation, described by the symplectic matrix
\cite{branature,network,3mj}
$$
T=\left(
\begin{array}{llllll}
 \frac{1}{\sqrt{3}} & 0 & \sqrt{\frac{2}{3}} & 0 & 0 & 0 \\
 0 & \frac{1}{\sqrt{3}} & 0 & \sqrt{\frac{2}{3}} & 0 & 0 \\
 \frac{1}{\sqrt{3}} & 0 & -\frac{1}{\sqrt{6}} & 0 & \frac{1}{\sqrt{2}} & 0 \\
 0 & \frac{1}{\sqrt{3}} & 0 & -\frac{1}{\sqrt{6}} & 0 & \frac{1}{\sqrt{2}} \\
 \frac{1}{\sqrt{3}} & 0 & -\frac{1}{\sqrt{6}} & 0 & -\frac{1}{\sqrt{2}} & 0 \\
 0 & \frac{1}{\sqrt{3}} & 0 & -\frac{1}{\sqrt{6}} & 0 & -\frac{1}{\sqrt{2}}
\end{array}
\right)\,,$$ the output covariance matrix of the three modes is
given precisely by
\begin{equation}\label{e:outmix}
 \gamma^{{\rm out}}=T \cdot [\gamma^{{\rm in}}_{1}(s,n)\oplus\gamma^{{\rm in}}_{2}(s,n)\oplus\gamma^{{\rm in}}_{3}(s,n)] \cdot T^T
 \equiv n \gamma (a)\,,
\end{equation}
where $\gamma(a)$ is defined in \eq{e:gamma} and we have made the
identification $a=[(s^2+2)/(3s)]$. \eq{e:outmix} describes
generally mixed, fully symmetric three-mode Gaussian states, with
global purity  given by $\mu=n^{-3}$, thus reducing to the pure
instance of \eq{e:gamma} for $n=\mu=1$. Recall that any additional
losses due e.g. to an imperfect tritter and/or to the distribution
and transmission of the three beams can be embedded into the
initial single-mode noise factor $n$, so that \eq{e:outmix}
provides a realistic description of the states produced in
experiments \cite{3mexp,naturusawa}. We therefore consider this
more general family of Gaussian states as resources to implement
the CV version of the primitive. It may be interesting to recall that
also for the general family of states of \eq{e:outmix} the genuine
tripartite entanglement is exactly computable \cite{3mj} in terms
of the residual contangle \cite{contangle}, and as expected, it
increases with $a$ and decreases with $n$.

Concerning the idealization (ii) discussed above, for an  efficient
implementation we should first consider the realistic case of nonideal homodyne detections, which means that the outcomes are affected by uncertainties quantified by the parameter $\sigma$, see \eq{overlap}. Furthermore, we should let the parties measure within a finite
range, which means specifically that in each measurement run the expected values for the measurement outcomes of the receivers can be shifted of some quantity $\Delta$ with respect to the corresponding expectation value of the sender's homodyne detection. So, what we should ask for is that the parties
(one sender $S$ and two receivers $R_0$ and $R_1$) agree on those outcomes of their measurement results for which
$|\hat x_S| =x_0 >0$ and $|\hat x_{R_0}| = |\hat x_{R_1}| = (x_0 +
\Delta) > 0$, once the sender has announced his or her output results $x_0$ to the other two parties. Here, the actual results of each measurement are assumed to distribute according to a Gaussian function centered at $x_0$ (for the sender) and $x_0+\Delta$ (for the receivers), respectively, with a variance   $\sigma$. Now the associated events occur with a finite probability. We
wish to investigate for which range of the positive shift
$\Delta \in [\Delta_{\min},\,\Delta_{\max}]$, being $\Delta_{\max}-\Delta_{\min}$ the range, the conditional probability $\tilde p$
still approaches $1/3$. The wider such range, the higher the
probability of success, or efficiency, of the protocol; of course the regions in the space of parameters such that the primitive can be efficiently implemented will depend on the specific boundaries
 $\Delta_{\min}$ and $\Delta_{\max}$ and not only on their difference.
 Notice that in order to ensure that the sign of the quadratures does not change when allowing a finite range of valid output values (and therefore, correlations between classical bits are properly extracted (see Sec. III)), we always demand the shift $\Delta$ to be positive. This condition can be further relaxed allowing for a non-symmetric range of permitted values around the broadcasted value $x_0$. This extra condition, which could be straightforwardly included in our expressions if one aims at maximizing the efficiency of our protocol, does not modify the analysis of the realistic
 implementation of the protocol we perform in this Section.
 For the sake of clarity, we remark once again the role of the parameter $\Delta$: Allowing a nonzero shift $\Delta$ in the computation of the probability $\tilde p$,  means considering the realistic case in which  the outcomes of the measurements performed by each receiver can be not coincident with each other, and not coincident even with the corresponding broadcasted value of $|x_0|$.  In what follows  we show that the protocol can be run successfully provided that the deviations $|x_{R_0}|-|x_{S}|$ and $|x_{R_1}|-|x_{S}|$ (in each run) both lie in between a $\Delta_{\min}$ and a $\Delta_{\max}$, which will be determined.

\begin{figure*}[t]
\includegraphics[width=17cm]{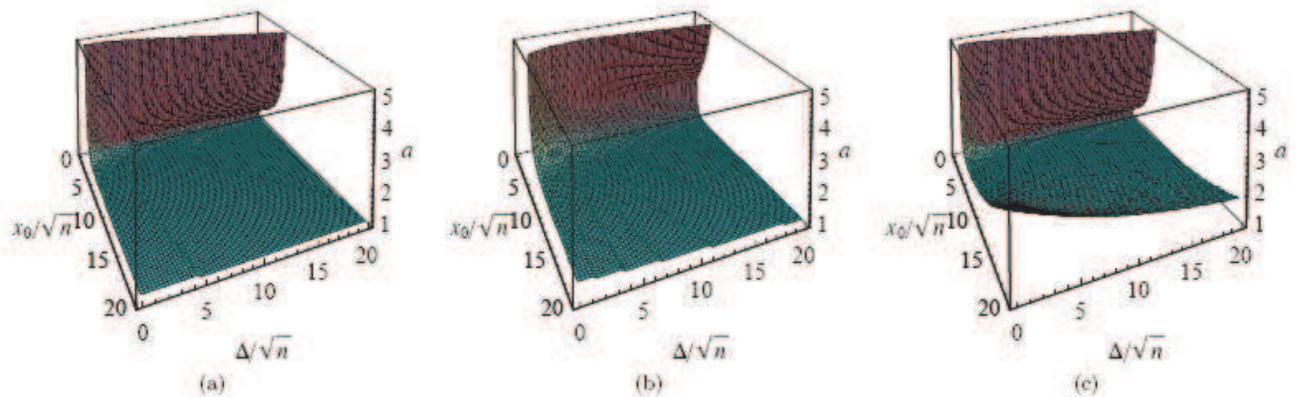}
 \caption{\label{parameters} (Color online) Plots, in the space of  $x_0/\sqrt{n}$
 (quadrature measurement normalized to noise), $\Delta/\sqrt{n}$
 (measurement shift normalized to noise), and $a$ (shared entanglement), of
 the boundary  to the useful region for which $\tilde p = 1/3 -
 \epsilon$ ($\epsilon=10^{-7}$), at different values of the homodyne detection uncertainty (a) $\sigma=0$ (ideal error-free  measurement), (b) $\sigma=1$ (fixed error, independent of the outcome), and (c) $\sigma=|x_0|/10$ (proportional error, corresponding to a $90\%$ efficiency in the detectors). Detectable broadcast can be
 solved efficiently, by means of our protocol, for all values of the parameters  which lie above the depicted surface. See text for an extended discussion. All of the quantities plotted are dimensionless.}
\end{figure*}

To this aim,
we can perform an analysis similar to the one of Section III, but
considering in full generality a non-unit noise factor $n$, a non-zero uncertainty $\sigma$, and a non-zero shift
$\Delta$. The calculations of the probabilities follow
straightforwardly along the lines of the special (unrealistic) case
previously discussed, and we finally obtain
\begin{eqnarray}\label{e:ptilde}
 \tilde{p}&=&\Bigg\{3+ 3 {\rm e}^{
-\frac{4 \left(\Delta +x_0\right) \left[\Delta  \left(4 \sigma ^2+7 a-3 R\right)+\left(4 \sigma ^2+3 a+R\right) x_0\right]}{3 \left(4 \sigma ^4+9 a \sigma ^2-R
   \sigma ^2+4\right)}} \nonumber \\
  & & +\ {\rm e}^{\frac{4 x_0 \left[4 (a-R) \Delta +\left(4 \sigma ^2+9 a-5 R\right) x_0\right]}{3 \left(4 \sigma ^4+9 a \sigma ^2-R \sigma ^2+4\right)}} \nonumber \\
 & & +\ {\rm e}^{-\frac{32 \left(\Delta +x_0\right) \left[\Delta  \left(\sigma ^2+a\right)+\left(\sigma ^2+R\right) x_0\right]}{3 \left(4 \sigma ^4+9 a \sigma ^2-R \sigma
   ^2+4\right)}}\,\Bigg\}^{-1}  \,,
\end{eqnarray}
with $R=\sqrt{9a^2-8}$. We find that in the parameter space of $a$
(regulating the entanglement), $n$ (regulating the mixedness),
$x_0$ (regulating the measurement outcome), $\sigma$ (regulating the measurement uncertainty), and $\Delta$
(regulating the measurement shift), there exists a surface which acts as the
boundary for the regime in which our primitive can be faithfully
implemented, yielding a feasible, robust solution to the broadcast
problem. This surface is obtained by requiring that $\tilde p = 1/3 - \epsilon$, with the deficit $\epsilon$ chosen arbitrarily
small. The result is plotted in Fig.~\ref{parameters} for $\epsilon=10^{-7}$ in the three-dimensional space of parameters
$x_0/\sqrt{n}$, $\Delta/\sqrt{n}$, and $a$, for different values of $\sigma$.
We consider the primitive as efficiently implementable in the whole region such that $1/3 - \epsilon\le \tilde p \le 1/3$, which spans the volume above the surface of Fig.~\ref{parameters}. Notice that for $\sigma=0$, $n=1$, and $\Delta_{\max} \rightarrow \Delta_{\min} \rightarrow 0$, such a useful volume shrinks to a two-dimensional slice (with $a \ge 5\sqrt2/6$) which represents the parameter range in which the ``ideal'' implementation described in Sec. III is successful.

The figure offers several reading keys. Let us investigate
independently how the possibility of solving detectable broadcast
via our protocol depends on the individual parameters. For
simplicity, we will consider $n$ fixed and eventually discuss its
role. We will also for the moment keep the idealization of error-free homodyne measurements ($\sigma=0$), corresponding to Fig.~\ref{parameters}(a): this makes the subsequent discussion more tractable. However we will relax such an assumption in the end to show that a realistic description of the homodyne measurement does not significantly affect the performance and the applicability of the protocol. This validates the claims of efficiency that we make in the following.

Let us henceforth start with the dependence of the solution  on $a$.
Somehow surprisingly, there exist lower {\em and upper} bounds on
the tripartite entanglement such that only for $a_{\min} \le a \le
a_{\max}$ the protocol achieves a solution. The bounds naturally
depend on $x_0$ and $\Delta$. Specifically, we observe that
$a_{\min}$ diverges for $\Delta=0$ and $x_0 \rightarrow 0$, meaning that no feasible
solution can be achieved in the low-$\Delta$, low-$x_0$ regime;
the reason being that near $x_0=0$ there is not possibility of associating
the classical bits "0" and "1" to positive/negative values of the quadrature.
The lower bound then goes down reducing to the already devised
threshold of $a_{thresh}\equiv5\sqrt2/6$ for $\Delta$ close to
zero and $x_0 \gg 0$, and eventually converges to $a=1$ (i.e. all entangled states are
useful) for any finite $x_0\gg 0$ and $\Delta \rightarrow \infty$. On the other hand,
the upper bound on $a$, which surprisingly rules out states with
too much entanglement, is obviously diverging at $\Delta=0$ but
becomes finite and relevant in the regime of small $x_0$ and large
$\Delta$, eventually reducing to $3/2$ for any finite $x_0$ and
$\Delta \rightarrow \infty$.

Summarizing, the two extremal regimes
corresponding to $\Delta=0$  on the one hand and $\Delta \rightarrow
\infty$  on the other hand, both allow a solution of detectable
broadcast via our protocol: The main difference is that in the
former case one needs states with an entanglement above $a_{\min}=5\sqrt2/6$,
while in the latter case one needs states whose entanglement is below
$a_{\max}=3/2$. The regime of finite shift $\Delta$ interpolates
between these two limits. This means, in terms of useful range,
and hence of efficient implementations, that if one is able to
produce entangled states precisely  with $5\sqrt2/6 < a < 3/2$,
the protocol is implementable with high efficiency for {\em any}
shift $\Delta \in [0,\infty)$ in the measurement outcomes. i.e.
with {\em an infinite range of variability} allowed for the
acceptable data resulting from homodyne detections performed by
the receivers, for a given outcome $x_0$ of the sender. This
information is important in view of practical implementations, and
becomes especially valuable since the engineering of the required
entangled resources appears feasible: A squeezing between $4.5$
and $6$ dB is required in each single mode, which is   currently
achieved in optical experiments \cite{7db9db}.

Fig.~\ref{parameters}(a), alternatively, shows that for a fixed
entangled resource, $a$, there exist minimum thresholds both for
$x_0$ and $\Delta$ in order to achieve a solution to detectable
broadcast. While the useful range for $x_0$ is always unbounded from
above, we find that, interestingly, an upper bound $\Delta_{\max}$
exists for $a>3/2$. Precisely, for a given $a>3/2$, with increasing
$x_0$ we observe that $\Delta_{\min}$ decreases and $\Delta_{\max}$
increases, i.e. the useful range spanned by $\Delta$
widens. Conversely, at a given $x_0$, $\Delta_{\max}$ decreases with
increasing entanglement $a$, reducing the parameter space in which a
solution can be found. Consistently with the previous discussion, we
conclude again that in realistic conditions ($\Delta > 0$) it is
better to have a moderate amount of shared entanglement to solve
broadcast with optimal chances.

Now, let us note that the noise parameter $n$ simply induces a
rescaling of both $x_0$ and $\Delta$, in such a way that with
increasing $n$ the surface bounding the useful range of parameters
shrinks, as it could be guessed (the noise degrades the performance
of the protocol). Still, with typical noise factors characterizing
experimental implementations, {\em e.g.} $n=2$, the protocol appears very
robust and the solution is still achievable, for a given amount of
entanglement (say $a \lesssim 3/2$), provided that $x_0$ and
$\Delta$ exceed $\sqrt2$ times their respective minimum thresholds
obtained for ideally pure resources ($n=1$).

\begin{figure*}[t]
\includegraphics[width=17cm]{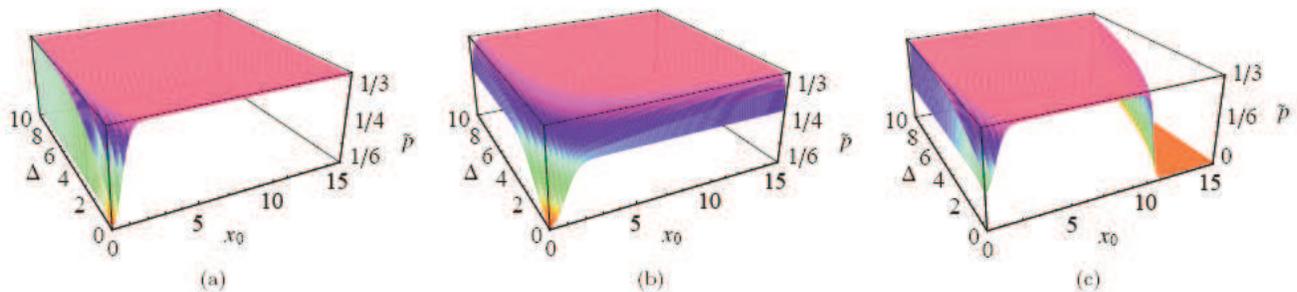}
 \caption{\label{plotpdn} (Color online) Plots of the conditional probability
 $\tilde{p}$,
 \eq{e:ptilde},
 as a function of the measurement outcome $x_0$ for the sender and the shift
 $\Delta$ of the measurement outcomes for the receivers, for realistically mixed Gaussian resource states with
 $n=2$ and
 $a=3/2$. The uncertainty in the homodyne detection is taken to be (a) $\sigma=0$, (b) $\sigma=1$, and (c) $\sigma=|x_0|/10$. Detectable broadcast is {\em efficiently} solvable in the wide
plateau region where
 $\tilde{p} \rightarrow 1/3$. All of the quantities plotted are dimensionless.}
\end{figure*}

Finally, let us address the important issue of the uncertainty affecting homodyne detections. Any realistic measurement is characterized by a nonzero $\sigma$. We have explicitly studied two situations, one in which the absolute error is fixed, corresponding to a constant $\sigma$ [see Fig.~\ref{parameters}(b)], and another in which the relative error is fixed, corresponding to a $\sigma$ proportional to the measurement outcome $x_0$ [see Fig.~\ref{parameters}(c)]. The result is that, in both cases, for not exceedingly high values of the error factor, the useful region is obviously reduced but, crucially, the possibility of achieving detectable broadcast via our protocol is still guaranteed in a broad range of values of the parameters. Specifically, as somehow expected,  the error model in which $\sigma$ is proportional to the measurement outcome results in a more consistent modification (shrinkage) of the useful surface, while almost nothing happens in the case of fixed, small $\sigma$.  In particular, upper bounds on $x_0$ arise for a practical realization in the presence of a proportional error, or in other words due to a limited efficiency in the detection. A significant portion of the parameter space anyway  remains valid for a workable implementation of the primitive. Our scheme is thus robust also with respect to the imperfections in the quadrature measurements. We draw the conclusion that the  protocol we designed is truly efficient and realistically implementable in non-ideal conditions.

Fig.~\ref{plotpdn} explicitly
depicts $\tilde{p}$, \eq{e:ptilde}, for sensible resource values of $a=3/2$ and $n=2$, as a function of $x_0$
and $\Delta$, and according to different values of $\sigma$ like in Fig.~\ref{parameters}. We notice a huge (unbounded from above in the ideal case $\sigma=0$) region in which
$\tilde{p}\rightarrow 1/3$, yielding an efficient solution to
detectable broadcast via our protocol. We have thus identified an
optimal ``work point'' in terms of the parameters describing the
shared Gaussian states, and we have defined the frontiers of application of our scheme in the laboratory practice. In this way, together with the
explicit steps of the protocol presented in the preceding sections,
we obtain a clear-cut recipe for a CV demonstration of detectable
broadcast, which we hope can be experimentally implemented in the
near future.

For the sake of completeness, let us mention that from
\eq{e:ptilde} we analytically find that $\tilde p$ converges {\em
exactly} to $1/3$ only in the following limiting cases (for a
given finite $n$ and $\sigma=0$): (i) $a \rightarrow \infty$ with $\Delta < 3x_0$;
(ii)  $x_0 \rightarrow \infty$ with $a \ge 5\sqrt2/6$; (iii)
$\Delta \rightarrow \infty$ with $1<a\le 3/2$.

\section{Concluding remarks}

In this paper we have proposed a protocol to  solve detectable
broadcast with entangled continuous variables  using Gaussian states
and Gaussian operations only. Our algorithm relies on genuine
multipartite entanglement distributed among the three parties, which
specifically must share two copies of a three-mode fully
symmetric Gaussian state. Interestingly, we have found that
nevertheless not all entangled symmetric Gaussian states can be used
to achieve a solution to detectable broadcast: A minimum threshold
exists on the required amount of multipartite entanglement. We have
moreover analyzed in detail the security of the protocol.

In its ideal
formulation, our protocol requires that the parties share pure
resource states, and that the outcomes of homodyne detections are
perfectly coincident and not affected by any uncertainty; this however entails that our protocol
achieves a solution with vanishing probability. To overcome such a
practical limitation, we have eventually considered a more realistic
situation in which the tripartite Gaussian resources are affected by
thermal noise, and, more importantly, the homodyne detections are realistically imperfect, and there is a finite range of
allowed values for the measurement outcomes obtained by the parties.
We have thoroughly investigated the possibility to solve detectable
broadcast via our protocol under these relaxed conditions. As a
result, we have demonstrated that there exists a broad region in the
space of the relevant parameters (noise, entanglement, measurement
range, measurement uncertainty) in which the protocol admits an efficient solution. This
region encompasses amounts of the required resources which appear
attainable with the current optical technology (with a legitimate
trade-off between squeezing and losses).
We can thus conclude that a
feasible, robust implementation of our protocol to solve detectable
broadcast with entangled Gaussian states may be in reach. This would
represent another important demonstration of the usefulness of
genuine multipartite continuous-variable entanglement for
communication tasks, coming to join the recent achievement of a
quantum teleportation network \cite{naturusawa}.

This work was initially supported by the Deutsche
Forschungsgemeinschaft, the Spanish Grant No. FIS-2005-01369,
CONSOLIDER-INGENIO Grant No. 2006-2010 CSD2006-0019 QOIT, Catalan Grant No.
SGR-00185, and EU IP Program SCALA.


\begin{thebibliography}{99}
\bibitem{notepurity} The
purity $\mu$ is defined, for a quantum state $\hat{\rho}$, as
$\mu={\rm Tr} \hat{\rho}^2$. If $\hat{\rho}$ denotes a Gaussian
state with covariance matrix $\gamma$, then $\mu=1/\sqrt{\det
\gamma}$.
\end{thebibliography}
\end{document}